\documentclass[aps,prl,twocolumn,
nofootinbib,tightenlines,showpacs,%
showkeys,floatfix,tightenlines,preprintnumbers]{revtex4}

\usepackage{epsfig,bm,feynmf}

\begin{document}



\title{The widths of quarkonia in quark gluon plasma}


\author{Yongjae Park}%
\email{sfy@yonsei.ac.kr}

\author{Kyung-Il Kim}%
\email{hellmare@yonsei.ac.kr}

\author{Taesoo Song}%
\email{songtsoo@yonsei.ac.kr}

\author{Su Houng Lee}%
\email{suhoung@phya.yonsei.ac.kr}

\affiliation{Institute of Physics and Applied Physics,
Yonsei University, Seoul 120-749, Korea}

\author{Cheuk-Yin Wong}%
\email{cyw@ornl.gov}

\affiliation{Physics Division, Oak Ridge National Laboratory,
Oak Rige, TN 37830}
\affiliation{Department of Physics, University of Tennessee,
Knoxville, TN 37996}


\begin{abstract}
Recent lattice calculations showed that heavy quarkonia will survive
beyond the phase transition temperature, and will dissolve at
different temperatures depending on the type of the quarkonium. In
this work, we calculate the thermal width of a quarkonium at finite
temperature before it dissolves into open heavy quarks. The input of
the calculation are the parton quarkonium dissociation cross section
to NLO in QCD, the quarkonium wave function in a
temperature-dependent potential from lattice QCD, and a thermal
distribution of partons with thermal masses.   We find that for the
$J/\psi$, the total thermal width above 1.4 $T_c$ becomes larger
than 100 to 250 MeV, depending on the effective thermal masses of
the quark and gluon, which we take between 400 to 600 MeV. Such a
width corresponds to an effective dissociation cross section by
gluons between 1.5 to 3.5 mb and by quarks 1 to 2 mb at 1.4 $T_c$.
However, at similar temperatures, we find a much smaller thermal
width and effective cross section for the $\Upsilon$.

\end{abstract}


\maketitle


\section{Introduction}

Recently, a number of important progress have been made in the
physics of $J/\psi$ suppression as a signature of quark gluon plasma
that inevitably leads us to augment the original work by Matsui and
Satz\cite{Matsui86} with a more detailed study of heavy quark system
at finite temperatures, before confronting the recent RHIC
data\cite{Phenix1}, and in predicting results for LHC. Among these
theoretical developments are the phenomenologically successful
statistical model for $J/\psi$
production\cite{Gorenstein99,PBM99,PBM06}, based on a coalescence
assumption near $T_c$\cite{PBM06}, the recombination of charm pairs
into $J/\psi$\cite{The06}, and the recent lattice calculations,
showing strong evidence that the heavy quarkonium will persist above
$T_c$\cite{Hatsuda03,Hatsuda04,Datta03,Datta05,Datta06}. While these
results seem at odd with each other, it only suggests that one still
needs a more detailed understanding of the properties of heavy quark
system in the quark gluon plasma, especially between the phase
transition and the dissolving temperatures, before a consistent
picture of quarkonium suppression in heavy ion collision can be
achieved.

In this respect, an  important quantity to investigate is the
effective thermal width, and/or the effective dissociation cross
section of a heavy quarkonium in the quark gluon plasma.  Except for
its existence, the present lattice results are far from making
quantitative statements on the magnitude of thermal width for
charmonium states above $T_c$.   Hence, in this work, we will use
the perturbative QCD approach to calculate the thermal width. So
far, such calculations have been limited to dissociation processes
by gluons to the lowest order
(LO)\cite{BP79,Lee87,KS94,AGGA01,Wong04,Blaschke:2005jg}, because
the elementary $J/\psi$-parton dissociation cross section was
available only to that order\cite{Peskin79}, and to part of the
next-to-leading-order (NLO) in the quasi free
approximation\cite{Rapp}. Recently, two of us have performed the
dissociation cross section calculation to NLO in QCD\cite{SL05}.
Here, we will implement the NLO formula, to calculate the
corresponding thermal width of charmonium\cite{Lee07}, and then
perform a similar calculation for the bottonium case.

The NLO calculation of $J/\psi$-parton dissociation calculations
involves collinear divergence.  When applying this elementary
cross section to dissociation by hadrons, the collinear divergence
is cured by mass factorization, which renormalizes the divergent
part of the cross section into the parton distribution function of
the hadron.   Such complications disappear at finite temperatures,
as the thermal masses of the partons automatically renders the
divergence finite.  The magnitude of the thermal mass as
a function of temperature has been obtained previously by examining
the equation of state \cite{Levai97}.  In the region of $T_c$ to
2$T_c$, they are of the order of 300-400 MeV for quarks and 400-600 MeV for gluons.
In this work, instead of following the detailed temperature dependence, we shall
study results with a thermal mass of 400 and 600 MeV for both the quarks and gluons.
 As we will see, with an effective thermal mass
of 400 MeV, we find that the effective thermal dissociation cross
section above 1.4 $T_c$ is larger than 250 MeV.  The NLO calculation
is proportional to the derivative of the momentum space wave
function.  We will use the Coulomb wave function, whose size have
been fitted to reproduce the result obtained by Wong\cite{Wong04},
using a potential extracted from lattice gauge thermodynamical
quantities.

In Section II, we will recapitulate the LO result.  In Section III,
we will discuss the NLO for $J/\psi$.  The LO and NLO results for
bottonium are given in Section IV.

\section{LO result}

The LO invariant matrix element for the $J/\psi$ dissociation
cross section by gluon first obtained by Peskin\cite{Peskin79} and
rederived by one of us\cite{OKL02} using the Bethe-Salpeter
equation  is given as,

\begin{figure}
\centerline{
\includegraphics[width=4cm]{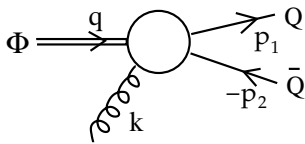}}
\caption{LO diagram} \label{lo-feynman}
\end{figure}

\begin{eqnarray}
\mathcal{M}^{\mu\nu} &=& -g \sqrt{\frac{M_\Phi}{N_c}} \left\{ {\bf
k} \cdot \frac{\partial \psi({\bf p})}{\partial {\bf p}}
\delta^{\nu 0} + k_0 \frac{\partial \psi({\bf p})}{\partial p^j}
\delta^{\nu j} \right\} \delta^{\mu i} \nonumber \\ && \mbox{}
\times \bar{u}(p_1) \frac{1+\gamma_0}{2} \gamma^i
\frac{1-\gamma_0}{2}T^a v(p_2)
\label{eq:amp}
\end{eqnarray}
Here, $\mu,\nu$ represents the polarization index of the $J/\psi$
and gluon respectively, and $k, p_1,p_2$ are the four momentum of
the gluon, $c, \bar{c}$. The quantity ${\bf p}$ is the relative
three momentum between $c$ and $\bar{c}$, and $\psi({\bf p})$ is the
charmonium wave function. $M_\Phi$ is the mass of a quarkonium, and
$N_c$ is the number of color.
Current conservation is easily shown to be
satisfied, $k_\nu M^{\mu \nu}=0$.
The energy conservation in the non-relativistic limit implies,
\begin{eqnarray}
k_0+m_\Phi=2m_c+\frac{{|\vec{p_1}|}^2+{|\vec{p_2}|}^2}{2m_c}
,
\end{eqnarray}
from which, the counting scheme for both
the LO\cite{Peskin79} and the  NLO\cite{SL05} are given as
follows,
\begin{eqnarray}
|\vec{p_1}|\sim|\vec{p_2}|\sim|\vec{p}|\sim O(mg^2) \nonumber \\
k^0 \sim |\vec{k}| \sim O(mg^4).
\end{eqnarray}

The effective thermal width and cross section are obtained by folding the matrix
element with the thermal parton distribution, $n(k_0)$
\begin{eqnarray}
\Gamma^{eff} & = &
d_p\int \frac{d^3k}{(2\pi)^3}n(k_0)v_{rel}\sigma(k_0), \nonumber \\
\sigma^{eff} & = & \int \frac{d^3k}{(2\pi)^3}n(k_0) \sigma(k_0)
 / \int \frac{d^3k}{(2\pi)^3}n(k_0), \label{g-lo}
\end{eqnarray}
where $d_p$ is the parton degeneracy, which is taken to be 16 for
the gluon in the LO calculation. Here, we will perform the
calculation in the rest frame of $J/\psi$, so the relative velocity
of $J/\psi$ and initial parton is $v_{rel}=|\vec{k}/k_{0}|$. The
cross section is given as,

\begin{eqnarray}
\sigma=\int\frac{1}{128\pi^2 m_\Phi
|\vec{k}|}\sqrt{\frac{k_0-\epsilon_0}{m_c}}
\overline{|\mathcal{M}|}^2 d\Omega\nonumber\\
\overline{|\mathcal{M}|}^2=\frac{2g^2m_c^2m_\Phi
(2k_0^2+m_{k_1}^2)}{3N_c} {\Big|\frac{\partial \psi({\bf
p})}{\partial {\bf p}}\Big|}^2 \label{lo-sigma}
\end{eqnarray}

\begin{figure}[h]
\centering \epsfig{file=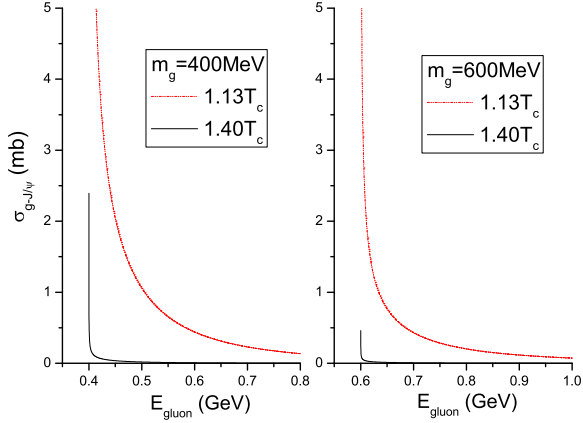, width=1.\hsize}
\caption{(Color online) $\sigma(E_{gluon})$ of $J/\psi$ at LO.}
\label{lo-diff-sig}
\end{figure}

\begin{figure}[h]
\centering \epsfig{file=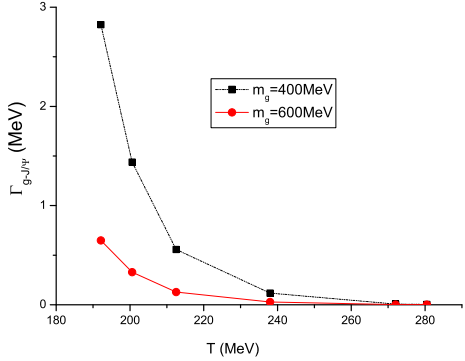, width=.8\hsize} \caption{(Color
online) Effective thermal width of $J/\psi$ at LO. The dots and
circles show the results at temperatures given in Table
\ref{jpsi-wave} with $T_c=170$MeV. } \label{width-lo}
\end{figure}

\begin{table}
\centering
\begin{tabular}{|c|c|c|c|c|c|c|}
\hline $T/T_c$ & 1.13 & 1.18  & 1.25  & 1.40 & 1.60 & 1.65 \\[2pt]
\hline $\epsilon_0$(MeV)& 36.4 & 20.9 & 10.1 & 3.4 & 0.14 & 0.004 \\[2pt]
\hline $\sqrt{<r^2>}$(fm) & 0.97 & 1.19 & 1.54 & 2.30 & 4.54 & 5.17 \\[2pt]
\hline $a_0$(fm) & 0.56 & 0.69 & 0.89 & 1.33 & 2.62 & 2.99 \\[2pt]
\hline
\end{tabular}
\caption{The binding energy, the rms radius, and its corresponding
Bhor radius of $J/\psi$ at finite temperature.} \label{jpsi-wave}
\end{table}

where $m_{k_1}$ is the thermal mass of a gluon. As can be seen from
Eq.(\ref{lo-sigma}), the cross section is proportional to the
absolute square of the derivative of the momentum space wave
function, which comes from the dipole nature of the interaction
between the gluon and the quark antiquark pair with opposite
charges. In the calculation of the  effective thermal width or the
effective cross section in Eq.(\ref{g-lo}), the cross section is
integrated over the incoming energy, which effectively integrates
over the absolute square of the derivative of the momentum space
wave function. As a consequence, the results are sensitive to the
size of the wave function only and not so much on its detailed
functional form. Therefore, we will use a Coulomb wave function,
whose Bohr radius is fitted to reproduce the rms radius obtained by
one of us\cite{Wong04} by solving the bound states in a
temperature-dependent potential extracted from lattice gauge
thermodynamical quantities.   For the binding energy, we use the
values obtained in ref. \cite{Wong04}. Table \ref{jpsi-wave}
summarizes the $J/\psi$ binding energy, its rms radius, and its
corresponding Bhor radius at finite temperature. The coupling
constant $g$ is set such that $\alpha_s$ is 0.5. As can be seen in
Fig. \ref{lo-diff-sig}, $\sigma(k_0)$ in the LO is dominant around
the $J/\psi$ binding energy\cite{Wong04}, which decreases as
temperature increases. On the other hand, $n(k_0)$ favors higher
$k_0$ as temperature increases.  Hence the overlap in
Eq.(\ref{g-lo}) decreases at higher temperature, more so because the
overlap integral starts from the thermal mass of the gluon. As can
be seen in Fig. \ref{width-lo}, choosing the effective thermal mass
of the gluons to be 400 and 600 MeV, which is much larger than the
binding energy, we find that the effective thermal width decreases
as the temperature increases and becomes very small. With lower
bound of the thermal mass (400 MeV) the width at LO is smaller than
3 MeV at 1.13 $T_c$ and less than 1 MeV at 1.4 $T_c$. At the NLO,
there are other thermal gluon and quark induced interaction
reactions representatively shown in Fig. \ref{NLO-graph}. The width
due to these contributions will be calculated in the next section.

\section{NLO result}

\begin{figure}[b]
\centerline{
\includegraphics[width=3.5cm]{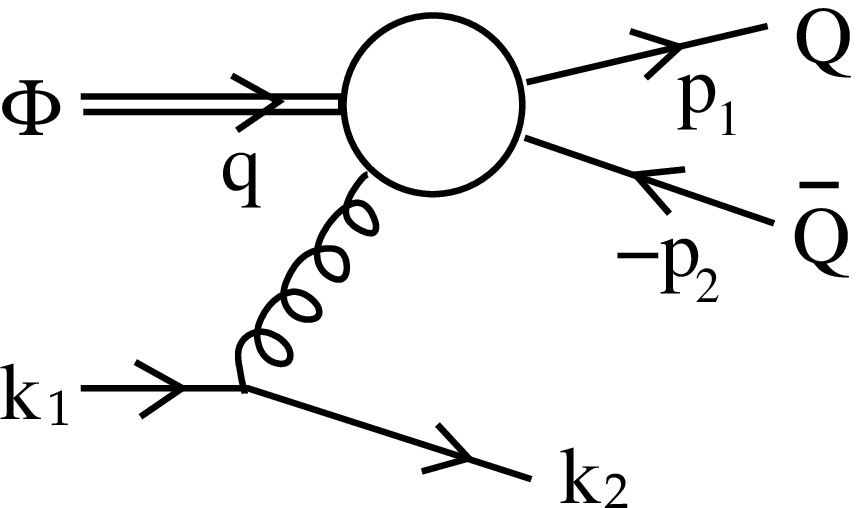}\hfill
\includegraphics[width=3.5cm]{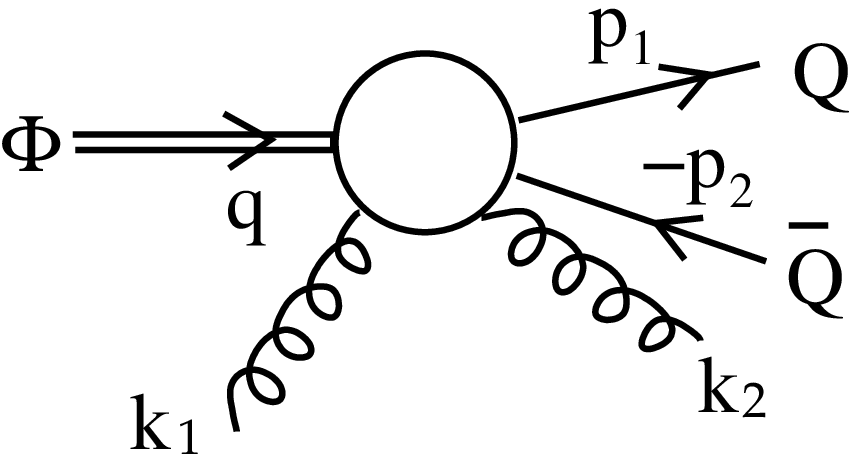}} \caption{NLO diagrams induced by (a)quarks,
(b)gluons.}\label{NLO-graph}
\end{figure}

The $J/\psi$ dissociation cross section by partons in QCD at the
NLO was performed by two of us\cite{SL05}.  The dissociation cross
section can be divided into two parts; the dissociation by quarks
and that by gluons (Fig. \ref{NLO-graph}).

\begin{figure}[h]
\centering \epsfig{file=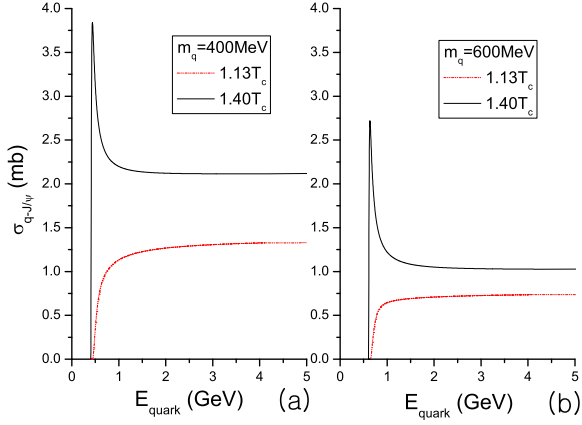, width=1.\hsize}
\caption{(Color online) $\sigma(E_{quark})$ of $q-J/\psi$ at NLO.}
\label{nlo-q-jpsi-diff-sig}
\end{figure}

\begin{figure}[h]
\centering \epsfig{file=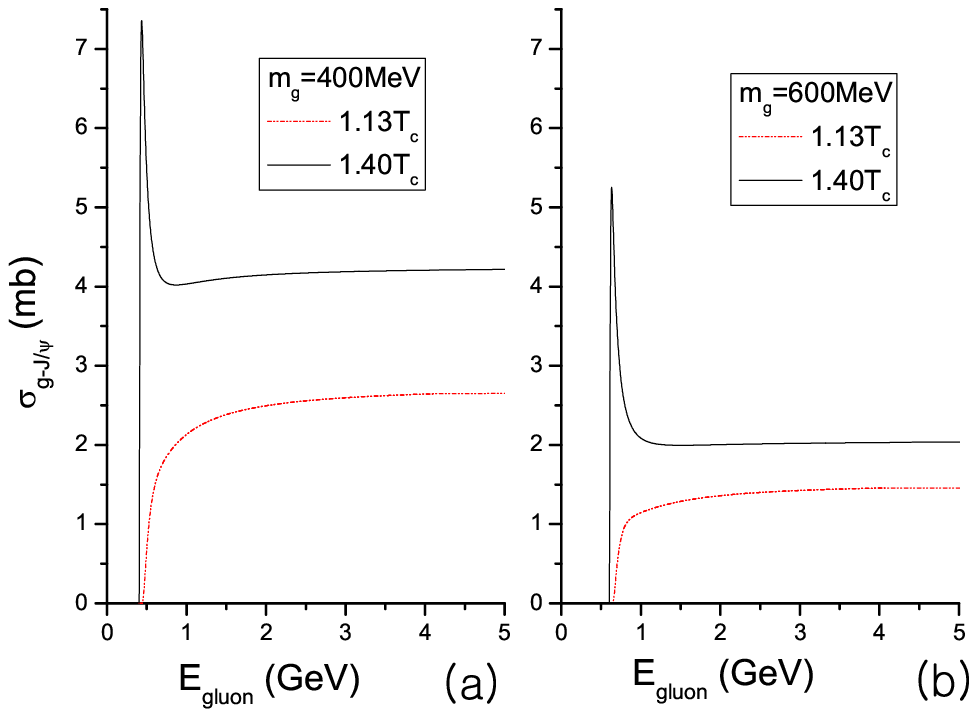, width=1.\hsize}
\caption{(Color online) $\sigma(E_{gluon})$ of $g-J/\psi$ at NLO.}
\label{nlo-g-jpsi-diff-sig}
\end{figure}

\begin{figure}[h]
\centering \epsfig{file=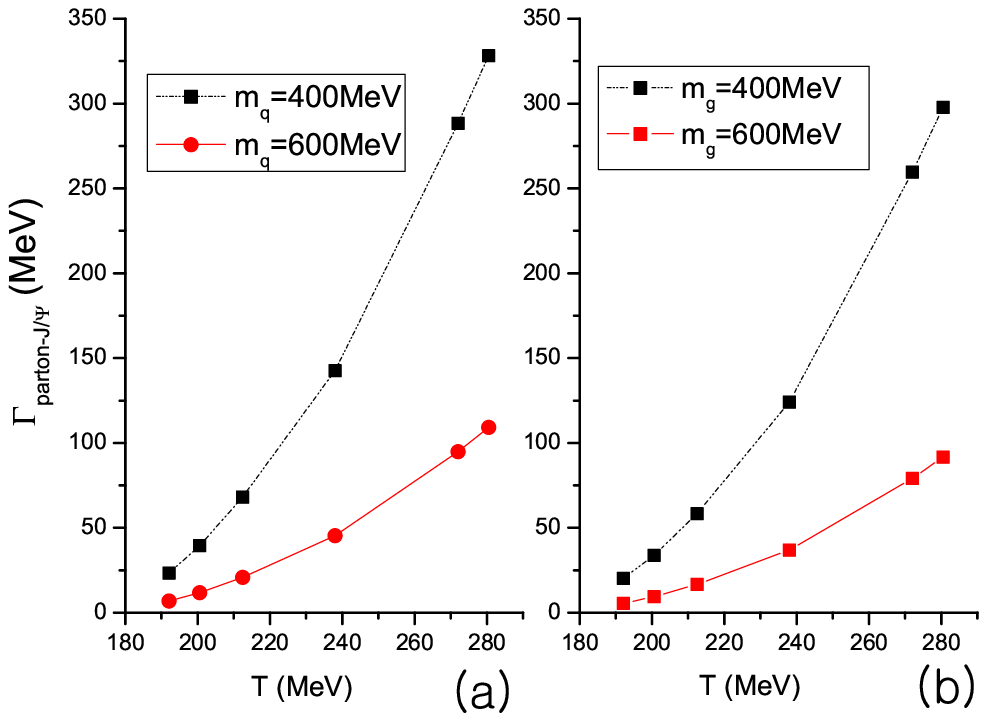, width=1.\hsize}
\caption{(Color online) Effective thermal width of $J/\psi$ at NLO
induced (a) by quarks and (b) by gluons} \label{jpsieffwidth}
\end{figure}

\begin{figure}[h]
\centering \epsfig{file=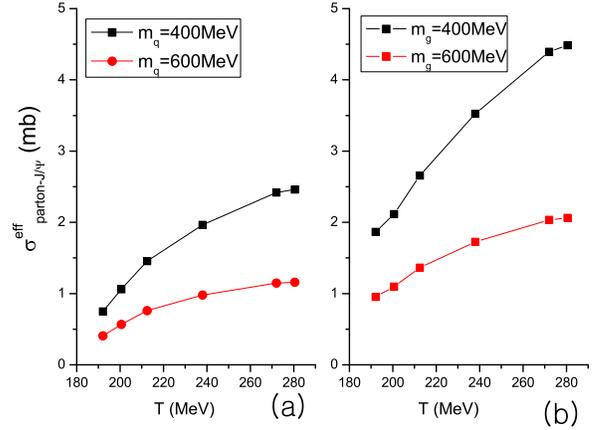, width=1.\hsize}
\caption{(Color online) Effective cross-section of $J/\psi$ induced
(a) by quarks (NLO) and (b) by gluons (LO+NLO)} \label{jpsi-effx}
\end{figure}

In both cases, the cross sections are given as,
\begin{eqnarray}
\sigma(u) = \frac{1}{4 \sqrt{(q \cdot k_{1})^{2}-m_{k_1}^2 m_\Phi^2
}}\int d\sigma_3 \overline{|\mathcal{M}|}^2
\end{eqnarray}
where $u^2=(q+k_1)^2$, $m_{k_1}$ is the thermal mass of a parton, and $d\sigma_3$ is the 3-body phase space,
\begin{eqnarray}
d\sigma_3 &=& \frac{d^3 k_2}{(2\pi)^3 2 k_{20}}
\frac{d^3p_1}{(2\pi)^32 p_{10}}\frac{d^3 p_2}{(2\pi)^3 2 p_{20}}
\nonumber\\&& \times (2\pi)^4 \delta^{(4)}(q+k_1-k_2-p_1-p_2)
\end{eqnarray}
Here, $q$, $k_1$, $k_2$, $p_1$ and $p_2$ are respectively the
momentum of $J/\psi$, incoming parton, outgoing parton, charm quark,
and anti-charm quark.

In the $J/\psi$ rest frame, the cross section can be written as,
\begin{eqnarray}
\sigma &=& \frac{1}{4 \sqrt{(q \cdot k_1)^2-m_\Phi^2 m_{k_1}^2}}
\nonumber
\\&& \times\int^{\beta}_{\alpha} {dw}^2 \int^{\beta'}_{\alpha'}
dp_\Delta^2 \frac{\sqrt{1-4 m_c^2/w^2}}{16^2 \pi^3 m_\Phi
|\vec{k_1}|} \overline{|\mathcal{M}|}^2,
\label{phase-space}
\end{eqnarray}
where $p_\Delta^2=(k_1-k_2)^2, w^2=(q+p_\Delta)^2$.  The
integration range is
\begin{eqnarray}
\alpha &=& (m_{p_2}+m_{p_1})^2=4m_c^2, \nonumber\\
\beta &=& (u-m_{k_1})^2, \nonumber\\
\alpha' &=& -b-\sqrt{b^2-ac}, \nonumber\\
\beta' &=& -b+\sqrt{b^2-ac},
\end{eqnarray}
where
\begin{eqnarray}
b &=& \frac{1}{2u^2}
\left\{u^2-(m_\Phi+m_{k_1})^2\right\}\left\{u^2-(m_\Phi-m_{k_1})^2\right\}\nonumber\\&&
-\frac{1}{2u^2}\left\{u^2-(m_\Phi^2-m_{k_1}^2)\right\}\left\{w^2-m_\Phi^2\right\},
\nonumber\\
b^2-ac &=&
\frac{1}{4u^4}\left\{(u^2-m_\Phi^2+m_{k_1}^2)^2-4u^2m_{k_1}^2\right\}\nonumber\\&&
\times\left\{w^2-(u+m_{k_1})^2\right\}\left\{w^2-(u-m_{k_1})^2\right\}.
\nonumber
\end{eqnarray}
The variables appearing in $\sigma$ can be expressed in terms of
$w^2, p_\Delta$ and $u^2$ as follows,
\begin{eqnarray}
q \cdot k_1 &=& (u^2-m_\Phi^2-m_{k_1}^2)/2, \nonumber\\
|\vec{k_1}| &=&
\sqrt{\left\{(u^2-m_{k_1}^2+m_\Phi^2)/(2m_\Phi)\right\}^2-u^2},
\nonumber\\
k_1 \cdot k_2 &=& (-p_\Delta^2+2m_{k_1}^2)/2, \nonumber\\
k_{10} &=& \sqrt{|\vec{k_1}|^2+m_{k_1}^2}, \nonumber\\
k_{20} &=& k_{10}-(w^2-p_\Delta^2-m_\Phi^2)/(2m_\Phi), \nonumber\\
|\vec{p}| &=& \sqrt{m_c(k_{10}-k_{20}+m_\Phi-2m_c)}.
\label{parameters}
\end{eqnarray}

We first consider the NLO effective thermal width induced by
quark. Here, the quark degeneracy is 36 assuming 3 flavors.  The
invariant matrix element is given as\cite{SL05},
\begin{eqnarray}
\overline{|\mathcal{M}|}^2=\frac{4}{3} g^4 m_c^2 m_\Phi
{\Big|\frac{\partial \psi({\bf p})}{\partial {\bf p}}\Big|}^2
\left(-\frac{1}{2}+\frac{k_{10}^2+k_{20}^2}{2 k_1 \cdot k_2}\right).
\label{m2-nloq}
\end{eqnarray}

Next, the gluon induced NLO calculation has the same degeneracy as
the LO case, and  the invariant matrix element is given as
follows\cite{SL05},
\begin{eqnarray}
&&\overline{|\mathcal{M}|}^2=\frac{4}{3} g^4 m_c^2 m_\Phi
{\Big|\frac{\partial \psi({\bf p})}{\partial {\bf p}}\Big|}^2
\Bigg\{-4+\frac{k_1 \cdot k_2}{k_{10}k_{20}}\nonumber \\&&
+\frac{2k_{10}}{k_{20}}+\frac{2k_{20}}{k_{10}}
-\frac{k_{20}^2}{k_{10}^2}-\frac{k_{10}^2}{k_{20}^2} +\frac{2}{k_1
\cdot k_2}\nonumber\\&& \times\left(
\frac{(k_{10}^2+k_{20}^2)^2}{k_{10}k_{20}} -2 k_{10}^2-2
k_{20}^2+k_{10}k_{20}\right) \Bigg\}.
\label{m2-nlog}
\end{eqnarray}

In the hadronic phase, the $1/(k_1\cdot k_2)$ term, and the
$1/k_{20}^2$ term give rise to collinear divergence, and soft
divergence respectively. However, in QGP phase, the thermal mass of
a parton $m_{k_1}$ in Eq.(\ref{parameters}) plays the role of a
cutoff. In contrast to the LO calculation, as can be seen from Fig.
\ref{nlo-q-jpsi-diff-sig} for quark and from Fig.
\ref{nlo-g-jpsi-diff-sig} for gluon, $\sigma(k_0)$ at the NLO does
not vanish at large $k_0$. This is so because irrespective of how
large the energies of the incoming quark or the gluon are, they can
always radiate small energy gluon in the order of the binding energy
to effectively dissociate $J/\psi$ via the LO process. Hence, in Eq.
(\ref{g-lo}), $\sigma$ has non trivial overlap with the maximum of
$n(k_0)$ that increase with temperature, leading to the result shown
in Fig. \ref{jpsieffwidth}(a) and in Fig. \ref{jpsieffwidth}(b) for
quark and gluon induced NLO width respectively. One thing to note
from Fig. \ref{nlo-q-jpsi-diff-sig} is that the elementary cross
section has a peak near threshold at 1.4 $T_c$. Such peak structure
only appears when the binding energy becomes very small and the
corresponding momentum space wave function becomes highly peaked
near zero momentum. When the incoming energy is small, these highly
peaked region gives important contributions to the two dimensional
phase space integral in Eq. (\ref{phase-space}). But when the
incoming energy becomes large, the phase space for the peaked region
becomes smaller and so does the total cross section. Such singular
behavior as a function of the incoming energy disappears when the
binding energy becomes larger.

Here we take the thermal mass from 400 MeV to 600 MeV\cite{Levai97}
within the temperature region of a few $T_c$. With those masses we
obtained large thermal widths. Even with an upper limit thermal mass
of 600 MeV, the width exceeds 100 MeV above 1.4 $T_c$, where we have
taken $T_c$=170 MeV. For example, if the thermal mass of partons is
600 MeV, and the produced $J/\psi$ remains at 1.4 $T_c$ for 2 fm/c,
its survival rate will be less than 40 $\%$.  As can be seen from
Fig. \ref{jpsi-effx}(a) and Fig. \ref{jpsi-effx}(b), with a 600 MeV
thermal mass at 1.4 $T_c$, the effective dissociation cross section
by a quark is about 1.0 mb and that by a gluon 1.5 mb.
 Hence, even though the $J/\psi$
might start forming at 1.6 $T_c$, its effective width is very large
and will not accumulate until the system cools down further.

\section{$\Upsilon$ Dissociation}

\begin{figure}[b]
\centering \epsfig{file=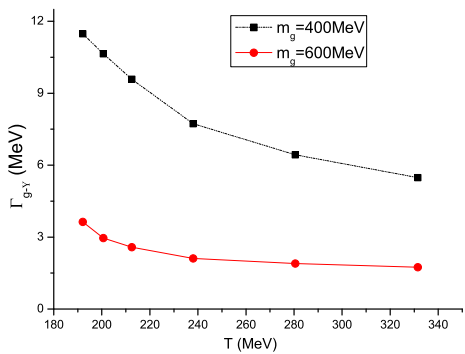, width=.8\hsize} \caption{(Color
online) Effective thermal width of $\Upsilon$ at LO}
\label{ups-width-lo}
\end{figure}

Here, we present the result for the $\Upsilon$ case. The $\Upsilon$
wave function is less sensitive to changes in the
temperature\cite{Wong04}.   In the LO calculation, while the trends
in the temperature dependence of $\sigma$ is similar to that of the
$J/\psi$, its variation in magnitudes is much smaller. Moreover, as
can be seen in Table \ref{upsilon-wave} the binding energy remains
large.  As a consequence, the overlap of $\sigma$ and the thermal
distribution remains large even at large temperatures, and the
effective thermal width slowly decreases with temperature as can be
seen in Fig. \ref{ups-width-lo}.

\begin{figure}[t]
\centering \epsfig{file=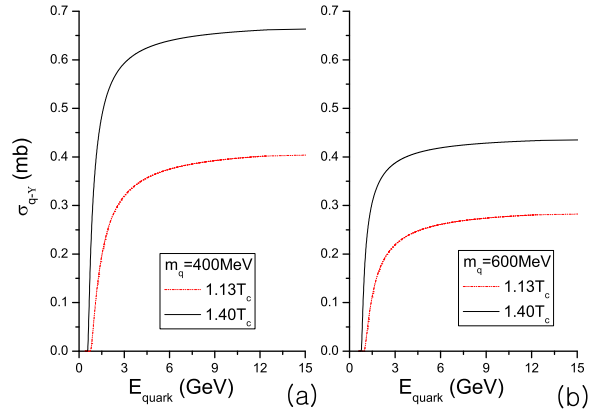, width=1.\hsize}
\caption{(Color online) $\sigma(E_{quark})$ of $q-\Upsilon$ at NLO.}
\label{nlo-q-ups-diff-sig}
\end{figure}

\begin{figure}[t]
\centering \epsfig{file=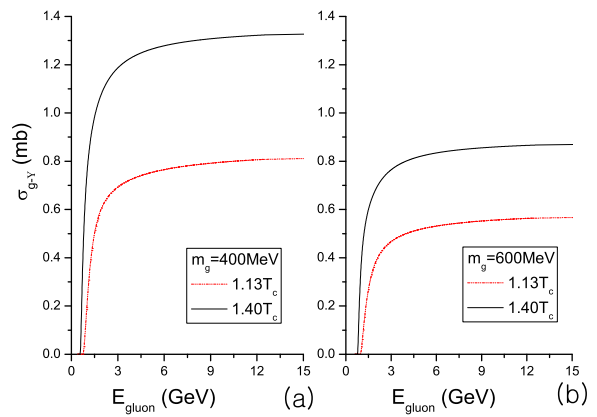, width=1.\hsize}
\caption{(Color online) $\sigma(E_{gluon})$ of $g-\Upsilon$ at NLO.}
\label{nlo-g-ups-diff-sig}
\end{figure}

\begin{figure}
\centering \epsfig{file=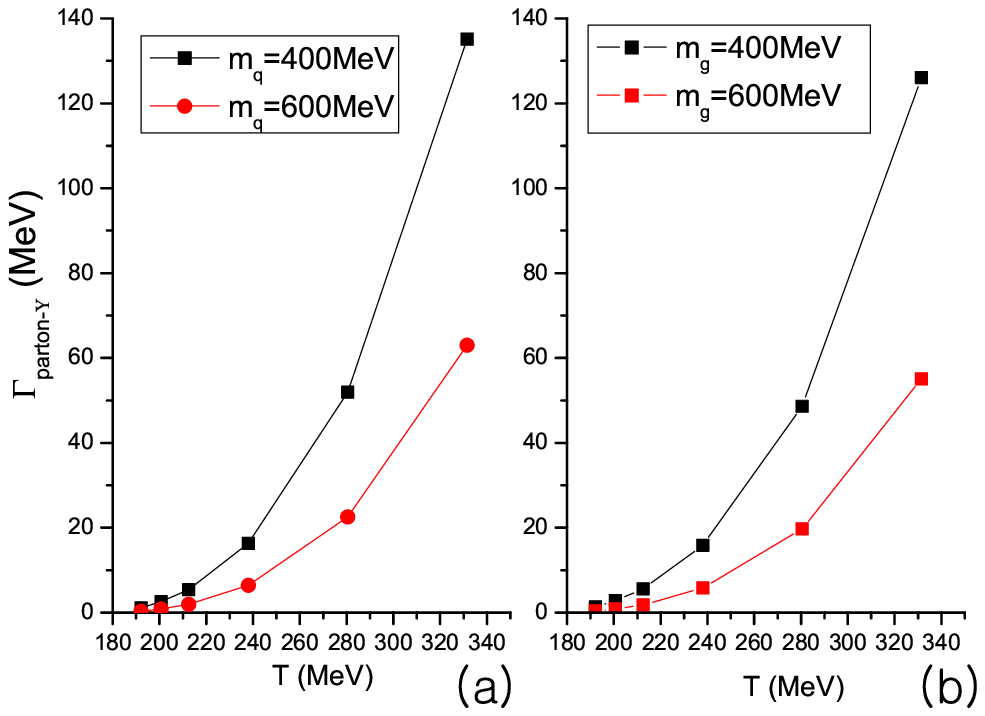, width=1.\hsize}
\caption{(Color online) Effective thermal width  of $\Upsilon$ at
NLO induced (a) by quarks and (b) by gluons.} \label{ups-width-nlo}
\end{figure}

\begin{figure}
\centering \epsfig{file=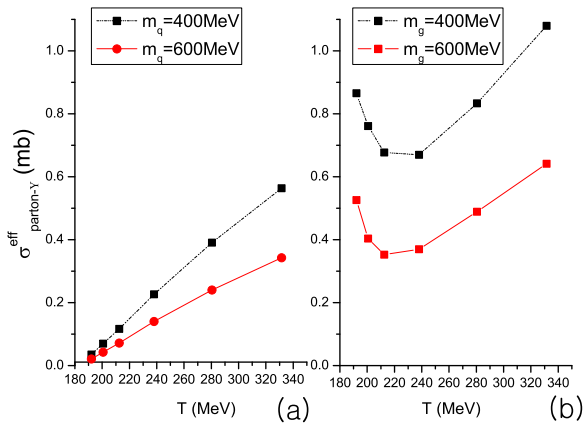, width=1.\hsize}
\caption{(Color online) Effective cross-section of $\Upsilon$ due to
(a) a quark and (b) a gluon.}\label{ups-effx}
\end{figure}

In the NLO calculation, as shown in Fig. \ref{nlo-q-ups-diff-sig},
and Fig. \ref{nlo-g-ups-diff-sig}, the form of the cross section is
similar to that of $J/\psi$.  However, because the binding energy of
$\Upsilon$ is still large at high temperature, there is no peak
structure near threshold.   Moreover, because $\Upsilon$ is more
tightly bound and has a smaller dipole size than $J/\psi$, its
corresponding $\sigma$ is also smaller. Therefore,  as can be seen
from Fig. \ref{ups-width-nlo}(a) for the quark induced and Fig.
\ref{ups-width-nlo}(b) for the gluon induced width, the overall
value of the width for $\Upsilon$ is smaller than that of $J/\psi$.
With an upper limit of thermal mass of 600 MeV and at temperature of
1.65 $T_c$, the sum of the LO and NLO thermal widths is less than 50
MeV. At this temperature, the effective dissociation cross section
by a quark is less than 0.2 mb and by a gluon than 0.6 mb.
Therefore, unlike the $J/\psi$ case, the $\Upsilon$ has a smaller
thermal width and effective dissociation cross section, and will
effectively start accumulating at higher temperatures. Fig.
\ref{ups-effx}(b) is the sum of LO and NLO effective cross section
of $\Upsilon$ due to gluon. Because the contribution of LO is not
negligible, the shape is quite different from the effective cross
section due to quarks of Fig \ref{ups-effx}(a).

\begin{table}[b]
\centering
\begin{tabular}{|c|c|c|c|c|c|c|}
\hline $T/T_c$ & 1.13 & 1.18 & 1.25 & 1.4 & 1.65 & 1.95 \\[2pt]
\hline $\epsilon_0$(MeV)& 313 & 247 & 203 & 150 & 111 & 86 \\[2pt]
\hline $\sqrt{<r^2>}$(fm) & 0.294 & 0.331 & 0.366 & 0.425 & 0.494 & 0.562 \\[2pt]
\hline $a_0$(fm) & 0.17 & 0.19 & 0.21 & 0.246 & 0.285 & 0.324  \\[2pt]
\hline
\end{tabular}
\caption{Binding energy and rms of the $\Upsilon$.}
\label{upsilon-wave}
\end{table}

\section{Summary}

In this work, we have calculated the thermal widths of $J/\psi$ and
$\Upsilon$ at finite temperature using the elementary
parton-quarkonium dissociation cross section at NLO in QCD and
assuming thermal partons with effective thermal mass.  We find that
for $J/\psi$ at 1.4 $T_c$, the thermal width will be 100 to 250 MeV,
which translates into an effective thermal cross section of several
mb.  However, the corrsponding width and effective cross section for
the  $\Upsilon$ is much smaller. Recently Mocsy and Petreczky
\cite{Mocsy:2007jz} have also calculated the thermal widths of
charmonium and bottonium in QGP, assuming that the quarkonium and
its constituents are in thermal equilibrium with the surrounding.
The thermal width estimated by MP is similar to ours for the
bottonium but several times larger for the charmonium. The result
for the charmonium by MP is obtained using a phenomenological
formula obtained when the binding energy is much smaller than the
temperature; hence further work has to be performed to understand
the discrepancy.

\section{Acknowledgement}

The work was supported by Korea Research Foundation KRF-2006-C00011
and in part by the US National Science Foundation under contract
number NSF-INT-0327497 with the University of Tennessee. The work of
SHL was also supported by the Yonsei University Research Grant. We
would like to thank C. M. Ko, and R. Rapp for useful discussion.

\end{document}